\documentclass[ams,preprint]{revtex4}
\usepackage{amsmath}
\usepackage{epsfig}

\newcommand{\be}{\begin{equation}}
\newcommand{\ee}{\end{equation}}
\newcommand{\s}{\cite}
\newcommand{\la}{\label}
\newcommand{\ber}{\begin{eqnarray}}
\newcommand{\eer}{\end{eqnarray}}
\newcommand{\nn}{\nonumber}

\begin{document}
\title{Acoustic wave front reversal in a three-phase media}
\newcommand{\mgu}{M.V. Lomonosov Moscow State University, Research Computing Center,\\  Vorobyovy Gory, 
Moscow 119991, Russia}
\author{N.I. Pushkina}
\email{ N.Pushkina@mererand.com}

\begin{abstract}
Acoustic wave front conjugation is studied in a sandy marine sediment that contains air bubbles 
in its fluid fraction. The considered phase conjugation is a four-wave nonlinear parametric 
sound interaction process caused by nonlinear  bubble oscillations which are known 
to be dominant in acoustic nonlinear interactions in three-phase marine sediments. 
Two various mechanisms of phase conjugation are studied. One of them is 
based on the stimulated Raman-type sound scattering on resonance bubble oscillations. 
The second one is associated with sound interactions with bubble oscillations 
which frequencies are far from resonance bubble frequencies. 
 Nonlinear equations to solve the wave-front conjugation problem are derived,  expressions for 
acoustic wave amplitudes with a reversed wave front are obtained and compared for various frequencies of 
the excited bubble oscillations.

\end{abstract}
\maketitle

\section{Introduction}
In the present paper two various phase conjugation processes are 
considered and compared: A. the process based on the stimaulated Raman-type acoustic scattering by 
natural bubbles oscillations; B. the process which involves bubble oscillations (or better  
say bubble gtating) either with zero frequency or bubble oscillations with double signal 
and pump waves frequency which does not coinside with a resonance bubble frequency. 
In all these cases two opposing pump waves with equal frequencies 
and a signal wave at an angle to them propagate in a sediment. If the difference 
frequency of the pump  and  signal waves is just equal to the natural frequency of bubble 
oscillations, there can take place nonlinear Raman-type scattering of the pump wave into the signal wave 
by these natural bubble oscillations. The same bubble oscillations can induce scattering of the 
second pump wave into the reversed signal wave only in case of  
much lower   resonance bubble frequency  than the frequencies of the pump waves 
in order the energy and momentum coservation 
laws be fulfilled for the interacting waves. The case B.  involves four sound waves 
with the same frequency and bubble "oscillations" 
either with zero frequency 
or bubble oscillations with double sound waves frequency. This case for media  
containing bubbles with high resonance frequencies can be more advantegeous 
for the wave front conjugation problem since  such media  are  known \s{p,zab} to manifest 
 significantly stronger acoustic nonlinear interactions than those with lower bubble resonance frequencies.
 Below these different cases are considered separately in detail.

\section{Theory}
We consider first the acoustic phase conjugation associated with stimalated Raman-type
 scattering of sound by nonlinear bubble oscillations.  
This process is analogous to  phase conjugation in nonlinear optics that  involves 
nonlinear light scattering by hypersonic waves, that is stimulated  Mandelstam-Brilluoin scattering 
(see Ref. \s {Zeld}).  
 Let the signal and conjugate waves 
propagate along the x-axis at an angle $\theta$ to the propagation direction of the 
pump waves. Raman-type stimulated scattering of sound by bubble oscillations in three-phase 
marine sediments 
was studied in Ref. \s{eng}. Dynamic equations obtained in this paper are of the form 
\ber
\frac{\partial^2p}{\partial t^2}-\frac{m}{\rho_{0f}G}\frac{\partial^2p}
{\partial x^2}-\frac{\nu}{\rho_{0s}G}\frac{\partial^2\rho_s}{\partial
 t^2}=\frac{mn}{G}\frac{\partial^2V}{\partial t^2}\nn\\
\frac{\partial^2\rho_s}{\partial t^2}(1-m)-
\frac{k+(4/3)\mu}{\rho_{0s}}\frac{\partial^2\rho_s}{\partial x^2}
 -\nu\frac{\partial^2p}{\partial x^2}=0, \la{1}
 \eer
where $p$ is the pressure in the water, $\rho_s$ and $\rho_f$ are the solid and liquid phase densities, 
 (the subscript  "0"  refers to the equilibrium values), $m$ is the porosity of a sediment, 
 $n$ is the bubble concentration and $V$ is the bubble volume, $\nu=1-m-k/k_s$,
 \[G=\frac{1-m}{k_s}+\frac{m}{k_f}-\frac{k}{k^{2}_{s}},\] 
where  $k_f$, $k_s$ and $k$ are the bulk moduli of the fluid, 
mineral grains constituting the frame, and of the frame itself;
$\mu$ is the shear modulus of the frame. 

Eqs.(\ref{1}) are to be supplemented with the nonlinear equation for an individual 
bubble motion \s{zab},
\be
\ddot V+\omega_0^2V+f\dot V-\alpha V^2-\beta(2\ddot VV+\dot V^2)+\mu V^3+\nu(V^2\ddot V+\dot V^2V)=
\epsilon p, \la{2}
\ee  
where $\omega_0$ is the resonance bubble frequency. The coefficients in Eq. (\ref{2})
are expressed through the equilibrium bubble volume $V_0$, its radius $R_0$ and 
the adiabatic index  $\gamma$,
\[\alpha =\omega_0^2(1+\gamma)/2V_0,\,\,\,\,\,\,\,\beta =1/6 V_0,\,\,\,\,\,\,\,\,
\mu=(\gamma+1)(\gamma+2)\omega_0^2/6V_0^2,\,\,\,\,\,\,\,\,\,
\nu=(2/9)V_0^2,\]
\[\epsilon =4\pi R_0/\rho_{0f},\,\,\,\,\,\,\,f=\delta\omega_0,\]
 $\delta$ is
 the dimensionless absorption coefficient of  bubble oscillations. The resonance bubble frequency $\omega_0$ 
 is related to the equilibrium pressure in the pore water and the bubble radius as 
\be 
 \omega_0^2=3\gamma P_0/\rho_{0f}R_0^2.   \la{11}
\ee
In  Eqs.(\ref{1}) nonlinear hydrodynamic terms are omitted since
nonlinear acoustic processes in porous media are  known to be governed mainly 
by  nonlinear oscillations of bubbles contained 
in the pore water \s{don, ch}. The only nonlinear term in Eqs. (\ref{1}) 
that ensures  the nonlinear process 
is the last term of the first equation which 
in fact  is the sum of the linear $V^l$ and the nonlinear part  $V^n$.
Let's denote  the amplitudes of the opposing pump waves as $P_1$ and $P_2$,  the amplitude 
of the signal wave $P_3$ and the conjugate wave amplitude  $P_4$. Since as it was noted in Introduction 
we suppose 
the resonance bubble frequency $\omega_0$  nonlinearly excited by the pump wave $P_1$ 
 and the signal wave $P_3$
 to be much less than the pump wave frequency $\omega$ 
we can put the signal-wave frequency $\omega_s \approx\omega$. The scattering of the  pump wave 
 $P_2$ on the same bubble oscillations gives the conjugate acoustic wave $P_4$ with practically the 
 same frequency as that of the signal wave. For such a situation using  Eqs.(\ref{1}),(\ref{2}) 
 we arrive at the  equations for the signal and conjugate waves
 \ber
\frac{dP_3}{dx}=-Ea_1(a_1|P_1|^2P_3+a_2P_1P_2P^*_4)\nn\\
\frac{dP_4}{dx}=-Ea_2(a_2|P_2|^2P_4+a_1P_1P_2P^*_3)  \la{3}
\eer
with 
\[E=\frac{\rho_fnc\epsilon^3}{L\cos\theta\omega(\omega^2-\omega{_0^2})^2\delta\omega{_0^2}},\,\,\,\,\,\,\,
a_1=\frac{\alpha-\beta(\omega^2+\omega{_0^2}+\omega\omega_0)}{\omega+2\omega_0},\,\,\,\,\,\,
a_2=\frac{\alpha-\beta(\omega^2+\omega{_0^2}-\omega\omega_0)}{\omega-2\omega_0},\]

\[L=1+\frac{\nu^2}{m}\frac{\rho_f}{\rho_s}K^{-1}\left(1+\frac{k+(4/3)\mu}{\rho_sc^2}K^{-1}\right),\,\,\,\,\,\,
K=1-m-
\frac{k+(4/3)\mu}{\rho_sc^2}.\] 

Eqs. (\ref{3}) are derived in the so called fixed-field approximation that permits to neglect 
the changes in the amplitudes of the pump waves  $P_1$ and  $P_2$ due to nonlinearity 
their changes being only because of linear bubble oscillations that cause dispersion in the medium 
which is not taken into account here. 

The solution to Eqs. (\ref{3}) for equal intensities of the pump waves is of the form 
\ber
\frac{P_4(0)}{P_3(0)}=\frac{a{_1}a{_2}\left(1-e^{-Al}\right)}
{\left(a{_1^2}+a{_2^2}e^{-Al}\right)}, \la{4}
\eer
where 
$A=E(a{_1^2}+a{_2^2})|P_{1,2}|^2$, $l$ is the interaction length.

It is seen from  (\ref{4}) that the conjugate wave amplitude can approach  the same order 
of magnitude as that of the signal wave if $e^{-Al}$ is not close to unity. For rather typical 
sediment parameters listed for instance in Refs. \s{tur,buch,med} the value of 
the coefficient $A$ can become  
large enough only for very  high bubble 
concentrations $nV \approx 10^{-1}$ and rather high pump-wave intensities, the frequencies 
being $\omega=2\pi\times10^4s^{-1},\,\,\, \omega_0=2\pi\times10^3s^{-1}$. With this we can conclude 
that wave-phase conjugation based on such Raman-type acoustic scattering is not favorable unlike 
a similar case of phase conjugation in nonlinear optics associated with Mandelstam-Brilluoin scattering.  
This result  is closely connected  with the fact noted in 
Introduction that nonlinear acoustic interactions are usually more pronounced for high-resonance 
bubble frequencies  (while in this case $\omega_0 \ll \omega$). This is reflected in Eq. (\ref{2}) 
where the nonlinear coefficients are inversely proportional to  the equilibrium bubble volume,
which can be expected since  for the same oscillation amplitudes 
 relative volume perturbations are  higher for smaller equilibrium volume values.  
Since the resonance bubble frequency 
is directly connected with a bubble radius through Eq. (\ref{11}), this frequency decisively 
influencies the nonlinear sound interaction.  In addition one can say that at a fixed  quantity $nV_0$  
a smaller equilibrium bubble volume $V_0$ means a higher 
bubble concentration $n$ that definitely affects the nonlinear process.

From this point of view it 
would be  preferable to use phase conjugation based on acoustic wave interactions   not with 
natural bubble oscillations, but with  induced bubble oscillations 
of appropriate frequencies (see also \s {zab})
to fulfil energy-momentum conservtion laws, at the same time resonance bubble 
frequencies being rather high. We shall study this type of 
phase conjugation in more detail. It splits into 
the following processes. As to the second order nonlinearity in Eq. (\ref{2})
 two nonlinear interactions can take place in this case.
The first one: the two pump waves of frequency $\omega$ generate bubble oscillations of  
frequency equal to $2\omega$ 
 with zero wave vector, and this oscillation scatters on the signal wave of frequency $\omega$ 
 into the conjugate wave.
The second process is somewhat different: the pump wave $P_1$ scatters on the signal wave 
into a bubble "oscillation" with zero frequency or better say into a bubble grating; and this bubble 
grating scatters the other pump way $P_2$ into the conjugate wave. As to the third order nonlinearity 
in Eq. (\ref{2}) this nonlinearity  mixes directly all the four waves with the same 
frequency $\omega$ generating  the conjugate wave. Taking into acount these three physical processes 
and using Eqs. (\ref{1}), (\ref{2}) we arrive at the equations for the amplitudes of the signal 
and conjugate  waves that are supposed to change slowly at a wavelength scale,
 \ber
\frac{dP_3}{dx}=-iF(a|P_1|^2P_3+bP_1P_2P^*_4)\nn\\
\frac{dP_4}{dx}=iF(a|P_2|^2P_4+bP_1P_2P^*_3).  \la{5}
\eer
 These equations are similar to Eqs. (\ref{3}), but with different coefficients,
\[F=\frac{\rho_fnc\omega\epsilon^3}{L\cos\theta\omega_0^2(\omega^2-\omega{_0^2})^4(\omega_0^2-4\omega^2)^2},
\]
\[a=(\alpha-\beta\omega^2)^2(\omega_0^2-4\omega^2),\,\,\,\,\,\,
b=(\alpha-\beta\omega^2)^2(\omega_0^2-4\omega^2)+(\alpha-3\beta\omega^2)^2\omega_0^2+
[(3/2)\mu-\nu\omega^2](\omega_0^2-4\omega^2)\omega_0^2.\]
Eliminating $P_3$ from Eqs. (\ref{5}) one gets the equation for $P_4$ as follows,
\be
\frac{dP^2_4}{dx^2}-2iFa|P_2|^2\frac{dP_4}{dx}-F^2(a^2-b^2)|P_1|^2|P_2|^2P_4=0.  \la{6}
\ee
Eq. (\ref{6}) can be simplified since in the approximation of a slowly varying amplitude 
the second derivative $d^2P_4/dx^2$ can be neglected, and the solution to Eq. (\ref{6}) is 
of the form 
\be
\frac{P_4(0)}{P_3(0)}=-\frac{2P_1P_2ab\left(1-e^{iBl}\right)}{|P_2|^2\left(a^2+b^2-2a^2e^{iBl}\right)} \la{7}
\ee
with $l$ being the interaction length and
\[B=\frac{1}{2}\frac{F}{a}(a^2-b^2)|P_1|^2.\]
Let us compare the effect of wave phase conjugation in two cases described by the relations  
(\ref{4}) and (\ref{7}). Examine them from the point of view of frequency dependence. In both 
cases $\omega$ is the frequency of the pump, signal and conjugate waves and  $\omega_0$ 
is the resonance bubble frequency. In case (\ref{4}) $\omega_0\ll\omega$, while in case (\ref{7})
the situation is just opposite, $\omega_0\gg\omega$.
The quantity $A$   determining essentially the phase conjugation 
in case  (\ref{4}) is proportional to $\omega(\omega_0/\omega)^4$  while in case (\ref{7}) 
the corresponding quantity $B$ depends on frequency as $\sim\omega$. This means that the effect of phase 
conjugation in Raman-scattering process is considerably less pronounced than in case 
of the induced bubble oscillations  with frequencies equal either to zero or  double   
frequency of the acoustic waves 
provided the resonance bubble frequency is much 
higher than the frequency of the sound waves. 
To perform a numerical estimate of the conjugate sound amplitude we shall use the following 
experimental data and the values of  sediment parameters \s {tur,buch,med},
$$\omega=2\pi\times10^4\,{\rm s}^{-1},\,\,\,\,\,\,\,\omega_0=2\pi\times10^5\,{\rm s}^{-1};\,\,\,\,\,\,\,
\rho_f=1\,{\rm g/cm}^3,\,\,\,\,\,\,\,\,\,\rho_s=2.65\,{\rm g/cm}^3, \,\,\,\,\,\,\,\,\,
m=0.4, \,\,\,\,\,\,\,\,$$
$$c\approx1.7\times10^5\,{\rm cm/s},\,\,\,\,\,\,\, 
k=10^9{\rm dyn/cm}^2,\,\,\,\,\,\,\,\,\,\,\mu=5\times10^8\,{\rm dyn/cm}^2,
\,\,\,\,\,\,\,\,\, k_s=3.6\times10^{11}\,{\rm dyn/cm}^2.$$
$$ \gamma=1.4,
\,\,\,\,\,\,\,\,\ nV_0=10^{-5},\,\,\,\,\,\,P_1=P_2\approx10^5\,{\rm dyn/cm}^2,\,\,\,\,\,\,\,\,
P_0\approx10^6\,{\rm dyn/cm}^2$$
For the acoustic-wave frequency $\omega\approx2\pi\times10^4 s^{-1}$ the amplitude damping coefficient 
$\alpha\sim4.0\times10^{-3} cm^{-1}$ \s{buch} which means that the effective 
interaction length 
is of the order of $(200-300) cm$. The numerical estimates show that for the 
parameters listed above  the conjugate wave amplitude $P_4$ can approach  the signal wave amplitude 
$P_3$ by the order 
of magnitude at a distance   within the attenuation length of the interacting acoustic waves.

\section{Conclusion}
The obtained results can be summerized as follows. Acoustic-wave phase conjugation 
based on nonlinear bubble oscillations contained in the pore water of marine sediments  
 is investigated. Two possible 
mechanisms of wave-front reversal are considered in detail. The first one is associated 
with Raman-type stimulated acoustic scattering on resonance-frequency bubble oscillations. 
The second one is based on the sound scattering by induced bubble oscillations which frequencies 
do not  coinside with a resonance bubble frequency of the sediment. In the case of
Raman-type sound scattering the natural frequency of bubble oscillations 
is to be much less than the acoustic-wave frequency in order energy-momentum 
conservation laws be fulfilled. While in the second case the resonance bubble frequency 
not involved directly in the scattering process can be much higher than the acoustic wave frequency. 
At the same time bubble oscillations with just high resonance frequencies are known to 
influence significantly nonlinear  acoustic interactions. The performed   
numerical estimates confirm this fact. In the second case   the amplitude of the reversed 
radiation can reach a measurable value for reasonable parameters of a sediment. While   
 in case of the Raman-type scattering the reversed radiation amplitude 
is several orders of magnitude less for the same sound frequencies and intensities.

\end{document}